\documentclass[twocolumn,trackchanges,twocolappendix]{aastex631}

\graphicspath{{Figures/}{../}}

\begin{document}

\title{Discovery of a Debris Disk Around TWA 20}



\author[0000-0001-5053-2660]{Skyler Palatnick}
\affiliation{Department of Physics, University of California, Santa Barbara, CA, 93106, USA \\}

\author[0000-0001-6205-9233]{Maxwell A. Millar-Blanchaer}
\affiliation{Department of Physics, University of California, Santa Barbara, CA, 93106, USA \\}

\author[0000-0002-2696-2406]{Jingwen Zhang}
\affiliation{Department of Physics, University of California, Santa Barbara, CA, 93106, USA \\}

\author[0000-0002-6964-8732]{Kellen Lawson}
\affiliation{Center for Space Sciences and Technology, University of Maryland, Baltimore County, 1000 Hilltop Circle, Baltimore, MD 21250, USA \\}
\affiliation{Astrophysics Science Division, NASA-GSFC, 8800 Greenbelt Rd, Greenbelt, MD 20771, USA \\}
\affiliation{Center for Research and Exploration in Space Science and Technology, NASA-GSFC, 8800 Greenbelt Rd, Greenbelt, MD 20771, USA \\}

\author[0000-0002-8984-4319]{Briley L. Lewis}
\affiliation{Department of Physics, University of California, Santa Barbara, CA, 93106, USA \\}

\author[0000-0003-4909-256X]{Katie A. Crotts}
\affiliation{Space Telescope Science Institute, 3700 San Martin Dr, Baltimore, MD 21218, USA \\}

\author[0000-0001-5365-4815]{Aarynn L. Carter}
\affiliation{Space Telescope Science Institute, 3700 San Martin Dr, Baltimore, MD 21218, USA \\}

\author[0000-0003-4614-7035]{Beth Biller}
\affiliation{Scottish Universities Physics Alliance, Institute for Astronomy, University of Edinburgh, Royal Observatory, Blackford Hill, Edinburgh, EH9 3HJ, UK}
\affiliation{Centre for Exoplanet Science, University of Edinburgh, Edinburgh, EH9 3HJ, UK}

\author[0000-0001-8627-0404]{Julien H. Girard}
\affiliation{Space Telescope Science Institute, 3700 San Martin Dr, Baltimore, MD 21218, USA \\}

\author[0000-0002-5352-2924]{Sebastian Marino}
\affiliation{Department of Physics and Astronomy, University of Exeter, Stocker Road, Exeter EX4 4QL, UK\\}

\author[0009-0000-0303-2145]{Rapha\"el Bendahan-West}
\affiliation{Department of Physics and Astronomy, University of Exeter, Stocker Road, Exeter EX4 4QL, UK\\}

\author[0000-0002-1652-420X]{Giovanni M. Strampelli}
\affiliation{Space Telescope Science Institute, 3700 San Martin Dr, Baltimore, MD 21218, USA \\}

\author[0009-0005-6943-6819]{Andrew D. James}
\affiliation{Department of Physics and Astronomy, University of Exeter, Stocker Road, Exeter EX4 4QL, UK\\}

\author[0009-0008-2252-7969]{Klaus Subbotina Stephenson}
\affiliation{Department of Physics and Astronomy University of Victoria 3800 Finnerty Road Elliot Building, Room 207. Victoria BC V8P 5C2\\}

\author[0000-0002-2683-2396]{Rodrigo Ferrer-Chavez}
\affiliation{Center for Interdisciplinary Exploration and Research in Astrophysics (CIERA) and Department of Physics and Astronomy, Northwestern University, Evanston, IL, 60208, USA}

\author[0000-0001-8568-6336]{Mark Booth}
\affiliation{UK Astronomy Technology Centre, Royal Observatory Edinburgh, Blackford Hill, Edinburgh EH9 3HJ, UK}

\author[0000-0002-9962-132X]{Ben J. Sutlieff}
\affiliation{Scottish Universities Physics Alliance, Institute for Astronomy, University of Edinburgh, Royal Observatory, Blackford Hill, Edinburgh, EH9 3HJ, UK}
\affiliation{Centre for Exoplanet Science, University of Edinburgh, Edinburgh, EH9 3HJ, UK}

\author[0000-0002-1838-4757]{Aniket Sanghi}
\affiliation{Cahill Center for Astronomy and Astrophysics, California Institute of Technology, 1200 E. California Boulevard, MC 249-17, Pasadena, CA 91125, USA}

\author[0000-0002-2428-9932]{Clémence Fontanive}
\affiliation{Trottier Institute for Research on Exoplanets, Université de Montréal, Montréal H3C 3J7, Québec, Canada\\}

\author[0000-0003-4203-9715]{Emily Rickman}
\affiliation{European Space Agency (ESA), ESA Office, Space Telescope Science Institute, 3700 San Martin Drive, Baltimore, MD 21218, USA\\}

\author[0000-0002-4388-6417]{Isabel Rebollido}
\affiliation{European Space Agency (ESA), European Space Astronomy Centre (ESAC), Camino Bajo del Castillo s/n, 28692 Villanueva de la Ca\~nada, Madrid, Spain\\}

\author[0000-0002-9803-8255]{Kielan Hoch}
\affiliation{Space Telescope Science Institute, 3700 San Martin Dr, Baltimore, MD 21218, USA \\}

\author[0000-0001-6396-8439]{William O. Balmer}
\affiliation{Department of Physics \& Astronomy, Johns Hopkins University, 3400 N. Charles Street, Baltimore, MD 21218, USA \\}
\affiliation{Space Telescope Science Institute, 3700 San Martin Drive, Baltimore, MD 21218, USA \\}

\begin{abstract}

We report the discovery of a debris disk surrounding the M3 star, TWA 20, revealed by JWST coronagraphic observations using the Near-infrared Camera (NIRCam). With reference-star differential imaging (RDI), we resolve the disk in scattered light in the F200W filter at a high signal-to-noise ratio and in the F444W filter at a low signal-to-noise ratio. The disk morphology and orientation are characterized via a forward modeling approach, where we determine a radius of 64.7$_{-6.5}^{+6.2}$ AU and an inclination of 70.1$^{\circ}$$_{-3.3}^{+2.5}$. Utilizing our forward model, we improve the fidelity of the debris disk image using model-constrained RDI (MCRDI). The newly discovered disk is one of only 6 disks detected in scattered light that orbit M dwarf stars; it is the third largest of the 6 resolved M dwarf disks and orbits the third faintest host star. The detection of this disk exemplifies the sensitivity of JWST to debris disks around low-luminosity host stars, which have historically been difficult to detect because these disks are cool and dim. We identify a nebulous structure that cannot be explained by an axisymmetric disk. A search for companions in the TWA 20 system yields no candidates. 

\end{abstract}


\section{Introduction} \label{sec:intro}

Extrasolar debris disks, much like our own Solar System's asteroid belt and Kuiper belt, can reveal information about planetary system formation, structure, and composition. Substructures within disks can also point to the presence of unseen planets \citep{ren2021layered}. As both tracers of hidden planets and windows into an earlier stage of planetary systems, debris disks play an essential role in aiding our understanding of the evolution of our Solar System and how it compares to the rapidly growing library of extrasolar systems. Comprised of remnant planetesimals and dust from their collisions, debris disks have been detected around stars with a range of different spectral types and characteristics \citep{hughes2018debris}. 

Thus far, most disks have been detected via the host star's spectral energy distribution (SED), where the thermal emission from the dust creates an ``infrared excess'' above the expected flux from the star at mid- and far-infrared (MIR/FIR) wavelengths. Infrared excesses were first detected by the Infrared Astronomical Satellite (\textit{IRAS}) survey \citep{aumann1984discovery,oudmaijer1992sao, mannings1998candidate}. The number of debris disk detections via infrared excess then increased by an order of magnitude with the launch of the \textit{Spitzer} \citep{su2006debris,trilling2008debris} and \textit{Herschel} surveys \citep{pilbratt2010herschel}; from these surveys, over 1000 stars with various spectral types and dozens of white dwarfs were found to have infrared excesses indicating that they host debris disks. Follow-up resolved imaging---with instruments such as the Space Telescope Imaging Spectrograph (STIS) and Near-infrared Camera and Multi Object Spectrometer (NICMOS) on the \textit{Hubble Space Telescope} (HST), SPectropolarimetric High contrast Exoplanet REsearch (SPHERE) instrument on the Very Large Telescope (VLT) \citep{beuzit2008sphere}, and the Gemini Planet Imager (GPI) on the Gemini South Telescope \citep{macintosh2006gemini}---proved to be important both for the validation of debris disk candidates identified by \textit{Spitzer} and \textit{Herschel} and characterization of the dust grains comprising the disks \citep{chen2020spitzer}. These high-contrast scattered light imaging observations, as well as interferometric measurements of thermal emission with the Atacama Large Millimeter/submillimeter Array (ALMA), also revealed a great variety of disk geometries; multiple rings, spiral arms, and warps are just a few of the many different configurations that have been observed to date. How exactly these features form and to what extent they are sculpted by unseen planets remains a clear gap in our knowledge of planetary system formation \citep{hughes2018debris}.
 
While \textit{IRAS}, \textit{Spitzer}, and other subsequent missions have identified debris disks around host stars of varied spectral types, M dwarf debris disk detections have remained notably scarce. This scarcity has been extensively documented in several previous works, detailing the lack of detection in the MIR, FIR, and submillimeter wavelength regimes \citep{plavchan2005m,gautier2007far,lestrade2009search,luppe2020MNRAS, lestrade2025debris}. Nonetheless, M dwarfs are still believed to host debris disks; observations of star-forming regions imply a similar occurrence rate of protoplanetary disks for M-type dwarfs compared to other spectral types \citep{andrews2005circumstellar}, which may suggest a similar occurrence rate of debris disks. Similarly, some survival models suggest that M dwarfs should host disks at rates similar to those of larger stars \citep{heng2013debris}. Given that M dwarfs make up the majority of stars in the Milky Way Galaxy \citep{miller1979initial} and as such likely account for a significant fraction of Galactic planetary systems, there is motivation to understand why M dwarf debris disks are so difficult to detect and to continue the search for disks and planets around M dwarfs. Several works have proposed theories about the scarcity of observed M dwarf debris disks which include the removal of detectable particles by stellar winds \citep{plavchan2005m,augereau2006microscopii,schuppler2015collisional} or coronal mass ejections \citep{osten2013coronal}, photoevaporation \citep{adams2004photoevaporation}, and the fact that M dwarfs are faint and prior surveys may not have had the sensitivity to detect their surrounding debris disks \citep{2014A&Alestrade,luppe2020MNRAS,cronin2023alma}. 

The M dwarf disks that have been detected and imaged exhibit a fascinating variety of morphologies and characteristics. The disk around AU Microscopii (AU Mic) has a two-ring structure \citep{strubbe2006dust,boccaletti2015fast}, substantial brightness asymmetries, and ripples that correspond to ``dust avalanches," or high-speed radial outflows of dust \citep{boccaletti2015fast,chiang2017stellar}, explained in part by stellar wind and in part by a potential planet orbiting within the disk \citep{plavchan2020planet}. The disk orbiting TWA 7 has three rings, a spiral arm, and a dust clump \citep{ren2021layered}. These features hint at the presence of at least one planet, and indeed a planetary candidate within the disk has recently been further validated \citep{lagrange2025evidence,crotts2025follow}.  

At the time of submission, five debris disks hosted by M dwarfs have been detected in scattered light. AU Mic was detected in scattered light via a coronagraphic instrument on the University of Hawaii 2.2m telescope \citep{kalas2004discovery}. Fomalhaut C (Fom C) was discovered via \textit{Herschel} \citep{kennedy2013discovery} and recently imaged in scattered light with JWST \citep{lawson2024jwst}. GSC 07396-00759 was discovered via VLT/SPHERE imaging, and intriguingly has no previously reported IR excess \citep{sissa2018new}. TWA 7 and TWA 25 were first imaged via HST/NICMOS \citep{choquet2016first}, though TWA 7 was initially detected via \textit{Spitzer} \citep{low2005exploring}. 

To enable more debris disk detections around M stars, high sensitivity observations are necessary. In the scattered light (visible to MIR) regime, JWST offers unprecedented sensitivity to faint debris disks, especially at long wavelengths inaccessible from the ground. As a space telescope, JWST is not limited by Earth's atmosphere, which is an obstacle for ground-based imaging at wavelengths greater than 2.1\textmu{m} \citep{rich2022gemini,ginski2024sphere,garufi2024sphere}. This is particularly useful for observing M dwarf disks, as these are red stars; light scattered by their surrounding dust may be brighter at near-IR and MIR wavelengths than at visible wavelengths \citep{lawson2024jwst}. JWST Near Infrared Camera (NIRCam), which operates between 0.6–5 \textmu{m}, is therefore uniquely equipped to detect such debris disks.

\begin{table*}
\caption{Stellar parameters for the TWA 20 system.}
\label{tab:twa20props}
\centering
\begin{tabularx}{\textwidth}{cccccc}
	    \hline
	    \hline
		Name & Spectral Type & V (mag) & Distance (pc) & Age (Myr) & $T_\mathrm{eff}$ (K) \\
            \hline
            TWA 20 & M3 & 13.24$\pm$0.13 & 80.21$\pm$0.02  & 10$\pm$2  & 3560$\pm$90 \\
            &\footnotesize\citep{shan2017multiplicity}&\footnotesize\citep{kiraga2012asas}&\footnotesize\citep{gaia2021}&\footnotesize\citep{luhman2023census}&\footnotesize\citep{weinberger2012distance}\\
		\hline
		\hline
	\end{tabularx}
\end{table*}

In this work, we describe the imaging discovery of a debris disk around TWA 20 using JWST NIRCam. The host is a faint \citep[V=13.24; ][]{kiraga2012asas} M3 star and has no previously detected IR excess. Stellar properties are summarized in Table \ref{tab:twa20props}. TWA 20 hosts the 6$^{\rm th}$ resolved debris disk around an M dwarf. TWA 20 was one of 23 science targets in the JWST General Observers (GO) program ``Uncharted Worlds: Towards a Legacy of Direct Imaging of Sub-Jupiter Mass Exoplanets" (GO 4050, PI: A. Carter), a survey designed to search for new sub-Jupiter mass companions in nearby systems in the TW Hydrae association \citep{crotts2025follow}. In the following sections, we report on TWA 20's disk properties, potential for planetary companions, and the potential for follow-up observations. 

\section{Observations} \label{sec:obs}

TWA 20 was observed with the JWST/NIRCam coronagraphy mode \citep{girard2022jwst} on June 14, 2024. The target was imaged using the \textbf{MASKRND} pupil plane mask and \textbf{MASK335R} focal plane mask with an inner working angle of $0\farcs63$. Observations were taken simultaneously in the F200W ($\lambda \sim 1.76-2.22 \ \mu$m) and the F444W ($\lambda \sim 3.89-4.91 \ \mu$m) filters.  The field of view (FOV) of the observations was $10\arcsec \times 10\arcsec$ for the F200W filter and $20\arcsec \times 20\arcsec$ for the F444W filter; the pixel scale was 31 mas per pixel for F200W and 63 mas per pixel for F444W. Images were taken at a telescope roll angle of 96.9$^{\circ}$ and the exposure time totaled 1860.13 seconds across 4 exposures for each filter. The M3 spectral type star HD 89063 \citep[V=9.34; ][]{hog2000tycho} was observed as a reference star for reference differential imaging (RDI), though ultimately all seven reference stars (HD 89063, i-Vel, HD 101581, IRAS 09432-4847, V* Y-Crt, IRAS 13041-5653, CD-23 9765) in the GO 4050 program were used in this data reduction. For each reference star, a 9-point grid dither pattern (step size either 15 or 20 mas depending on position) was used to compensate for any offset between TWA 20 and the center of the coronagraphic mask. Exposure times for the reference star observations were chosen to give a similar signal-to-noise ratio (SNR) to the science image each reference star corresponds to. 

\section{Data Reduction} \label{sec:drp}

We processed the science and reference images using a combination of the \texttt{spaceKLIP} \citep{Kammerer22, Carter23} and \texttt{Winnie} \citep{Lawson22} Python packages. The precise details of the \texttt{spaceKLIP} reduction can be found in \cite{Carter23} and are briefly summarized here. Once the files are read into the \texttt{spaceKLIP} pipeline, detector level corrections and ramp fitting are applied. Next, we complete background subtraction and flat-fielding, followed by median subtraction, bad pixel removal, and point-spread-function (PSF) centering and alignment. For the bad pixel removal step, an additional custom bad pixel map is constructed using the \texttt{ibadpx} software \citep{ibadpx} to flag any bad pixels missed during \texttt{spaceKLIP} detection. Zero additional bad pixels were identified with \texttt{ibadpx} for the F200W images, but 22 additional bad pixels were identified and removed in the F444W images. 

Following data pre-processing with \texttt{spaceKLIP}, we constructed and subtracted a model PSF to remove the stellar PSF from the science images. The PSF model was constructed from observations of each reference star in the GO 4050 survey. Reference stars were verified to ensure they do not have bright disks and are not binaries. Compared to a single reference star, a large catalog of reference stars, such as that from the full GO 4050 survey, results in a higher-fidelity PSF model \citep{crotts2025follow}. All but one reference star used here are M-type, and all but one reference star observations took place within a few days time span (as did the science observation), implying little wavefront drift. 
From the reference frames, we used \texttt{Winnie} to assemble the model PSF as a linear combination of each image based on the total intensity of the science frame \citep{Lawson22}. Once assembled, we used \texttt{Winnie} to carry out RDI. 

\begin{figure*}[!ht]
    \centering
    \includegraphics[width=\textwidth]{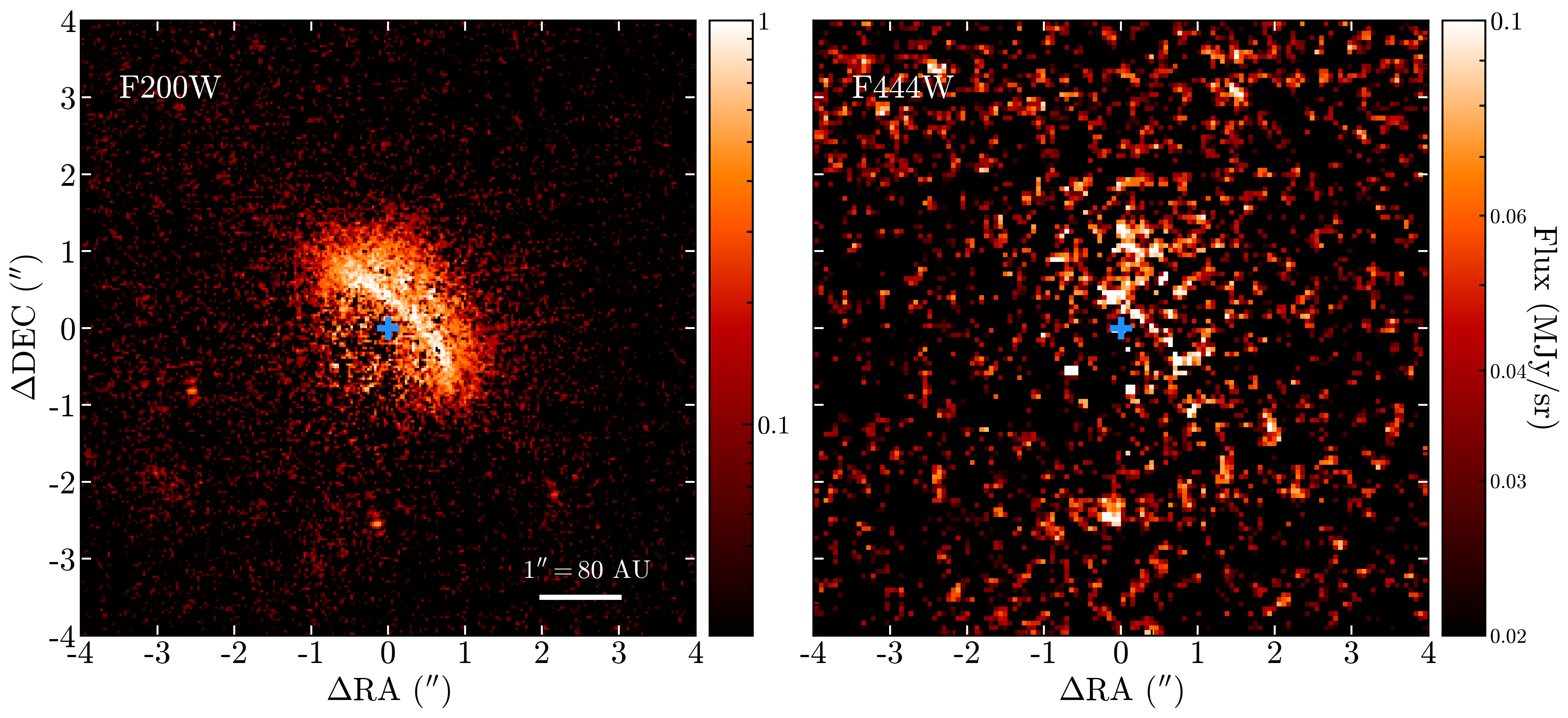}
    \caption{\label{fig:mcreduc} Final MCRDI data reductions for TWA 20 in the F200W (\textit{left}) and F444W (\textit{right}) filters. The images are rotated North up, and the $+$ represents the star location. Gaussian smoothing is applied to each image. Each tick represents 1$\arcsec$ ($\sim$80 au). The flux is log scaled in these images.} 
\end{figure*}

With conventional RDI, parts of the disk may be over-subtracted as a consequence of modeling the PSF of a science image with both stellar and circumstellar flux. As such, we employed model-constrained RDI (MCRDI) to improve the PSF subtraction and thus generate a higher SNR disk image  \citep{Lawson22}. To implement MCRDI for a given filter, we first used the RDI image output by \texttt{Winnie} to construct a forward model of the disk. Our forward modeling approach is described in Section \ref{sec:model}. \texttt{Winnie} regenerates a model PSF based on the total intensity of the disk-model subtracted science image and then repeats RDI with the new model PSF and the original pre-processed science image. Fig. \ref{fig:mcreduc} depicts the MCRDI images for both filters. We apply a Gaussian filter ($\sigma=0.5$ pixels) to smooth the image for visualization purposes. Measuring the average flux in the brightest region of the disk and comparing to the standard deviation of a featureless region of the image for each filter, we measure a 13$-\sigma$ disk detection in the F200W image and a 2$-\sigma$ disk detection in the F444W image.

In the F200W image of Fig. \ref{fig:mcreduc}, there are multiple off-axis point sources that are visible. Two of the three in this filter, as well as several more in the full F444W image, are introduced by the reference images and are thus spurious sources in the processed TWA 20 science images. As such, to conduct a search for companions in the system (Section \ref{sec:comp}), we re-ran the data processing with all off-axis sources in each reference image masked to prevent inclusion in the final reduction. 

\section{Disk Modeling}
\label{sec:model}
To execute MCRDI, quantify the geometry of the debris disk, and maximize our ability to find close-in companions, we employed a forward modeling approach with \texttt{Winnie}. This approach uses an initial guess of the disk parameters to generate a simple 2D disk model, convolves the disk model with the JWST - NIRCam 2.0\textmu{m} and 4.4\textmu{m} coronagraphic PSFs at their respective wavelengths, and then employs a minimization routine on the convolved model to find the best fit between the model disk parameters and the data. Additionally, for each iteration, we multiplied the raw model by a coronagraphic transmission map and then convolved each pixel with the nearest PSF spatial sample from a grid of PSFs sampled across the detector using the nearest optical-path difference (OPD) map, which is a wavefront error map estimated for the primary mirror. The OPD map must be accounted for as it can degrade the quality of the PSF. Our PSF grid was initialized with \texttt{STPSF} \citep{perrin2014updated}, which uses a model of JWST NIRCam, as well as a map defining OPD as a function of spatial position in a given image, to compute a simulated PSF.  The OPD map used in this reduction was measured on June 15, 2024, $\sim12$ hours after the science observation. The PSF grid sampled the synthetic PSF at 12 logarithmically spaced radial positions between $0.05-3.5\arcsec$ and 4 linearly spaced azimuthal positions between $0-360^{\circ}$ as well as one sample at the origin. The size of each grid axis was set to be 201 pixels oversampled by 2, such that the final generated PSFs used with Winnie were comprised of $402\times402$ points.

\begin{figure*}[!ht]
    \centering
    \includegraphics[width=\textwidth]{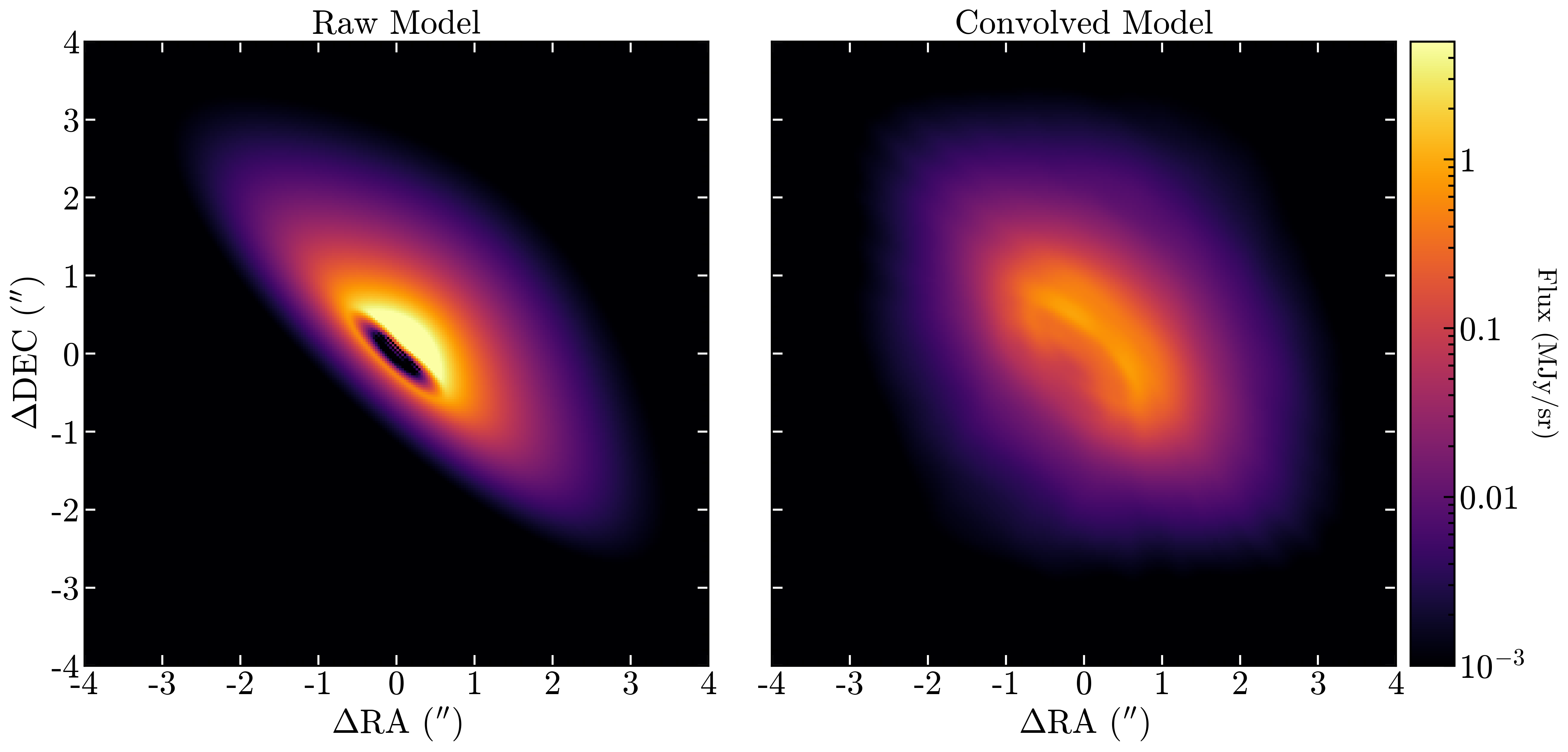}
    \caption{\label{fig:models} Optimized analytical disk model \textit{(left)} and the optimized convolved disk model \textit{(right)} constructed via forward modeling the data in the F200W filter.} 
\end{figure*}

We utilized a GRaTer disk model \citep{Augereau99} with the following free parameters: inclination ($\theta_{inc}$), position angle ($\theta_{PA}$), fiducial disk radius ($r_{0}$), the ratio between the scale height and $r_{0}$ at $r_{0}$ ($h_{0}$), the radial density power law exponent within $r_{0}$ ($\alpha_{in}$),  the radial density power law exponent beyond $r_{0}$ ($\alpha_{out}$), two Henyey-Greenstein asymmetry parameters ($g_{1}$ and $g_{2}$, respectively) \citep{HG41}, and the weight for the scattering phase function (SPF) with the asymmetry parameter $g_{1}$ ($w_{1}$). We also include $F_{max}$ as a free parameter, which scales the maximum brightness of the disk model to match the peak flux in the science image. Additionally, to slightly reduce computational complexity, we assume that the disk flaring parameter $\beta$ is 1 and the exponent of the vertical profile \textbf{$\gamma$} is 2. We chose to fix $\beta$ because values for $\beta$ are relatively consistently near unity across various disk observations \citep{williams2006dust,pawellek2014disk}; we chose to fix $\gamma$ because it is generally unconstrained in many debris disk images and modeling efforts \citep{lewis2025grater}. We choose our initial parameter values such that the input model is a visual match to the true disk image in terms of orientation and brightness; so long as these values provide a rough match to the data, the minimization routine will hone them to more accurate values. We find the best-fitting parameter values for each of the listed parameters using a Powell minimization routine from the \texttt{LMfit} Python package \citep{lmfit25}. The Powell algorithm represents a good compromise between computational speed and model accuracy \citep{varelas2019benchmarking}. For the F200W filter, the minimization routine converged after 954 model evaluations. Fig. \ref{fig:models} depicts the output of the modeling process for the F200W filter. The best fit parameter values from F200W modeling result were also used to create a forward model for the F444W filter.


\begin{figure*}[!ht]
    \centering
    \includegraphics[width=\textwidth]{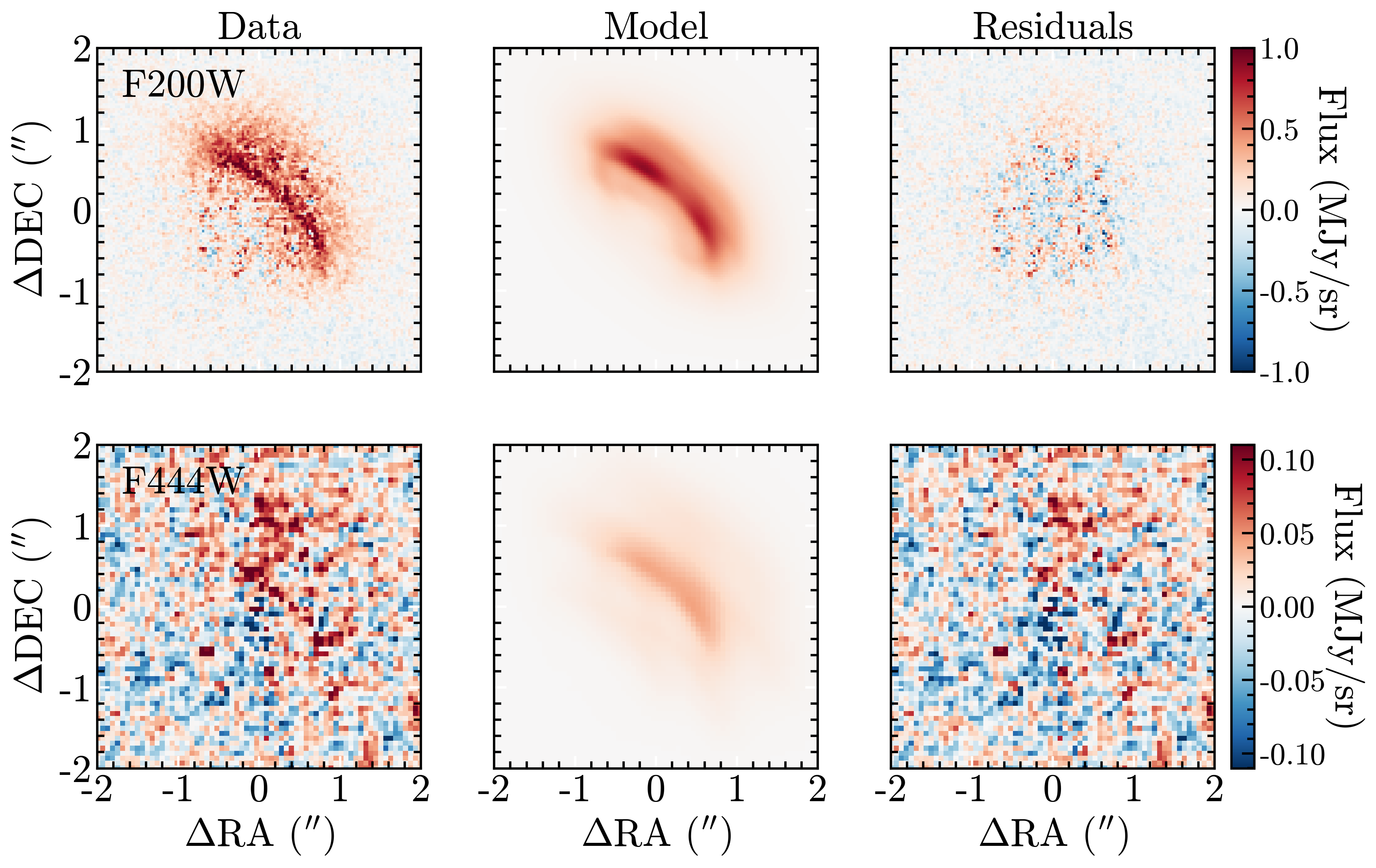}
    \caption{\label{fig:modelsandresids} Comparison of the forwarded modeled disk and MCRDI image in each filter. Residuals in the F200W filter show some slight over-subtraction, indicating potential over-fitting of the disk to the data. Given how faint the disk is in the F444W image, it is unclear how good of a fit the model is to the data. The smoothness of the model compared to the data is simply because the model is constructed (and displayed) without any noise, while the data has various noise sources.} 
\end{figure*}

Fig. \ref{fig:modelsandresids} compares the models in each filter with the corresponding MCRDI image. The F200W residuals show some negative disk structure, implying a slight oversubtraction. The oversubtraction is more substantial towards the southwest, which may indicate an asymmetry in the disk structure. It is also possible that for the structure of this particular disk, our two power-law model ($\alpha_{in}$, $\alpha_{out}$) is not adequate or that the SPF may vary with radius due to different grain size distributions. The disk emission is easily subtracted from the F444W image without leaving significant residuals. However, this filter provides very weak constraints to the model parameters given the low SNR. 

In order to more accurately constrain the debris disk physical parameters, we employed an additional nested-sampling disk fitting procedure to the F200W data using the \texttt{DYNESTY} Python software \citep{2020MNRAS.493.3132S,sergey_koposov_2024_12537467}. Described in detail in \cite{2004AIPC..735..395S,skilling2006nested}, nested-sampling optimization explores input parameter space by iteratively drawing a random set of sample points from the prior likelihood distribution for each parameter, constructing a model from these points, and computing a Bayesian likelihood \citep{akaike1998likelihood} from the residual difference between data and model. At each iteration, the prior distributions are constrained based on the sample with the lowest likelihood in the previous iteration, and new samples are drawn from these modified priors. Over many iterations, the lower-likelihood samples will be replaced by higher-likelihood samples, and the algorithm will ultimately terminate when the change in likelihood from one iteration to the next falls below a threshold value. We executed the nested sampling procedure with the same 2D GRaTer model parameters as the initial minimization. For each variable, we used a uniform prior spanning values above and below the forward-modeling best fit values with a minimum range of $\pm$40\% of the best fit value to ensure adequate exploration of parameter space. We determined uncertainties for each disk parameter from the set of posterior samples used in the optimization. Table \ref{tab:fitparams} summarizes the final debris disk parameter values derived from the two fitting procedures; 68\% confidence level uncertainties are included for any parameter fit with nested sampling. 

\begin{table}
\caption{Best fit geometric and scattering parameters for the TWA 20 debris disk. All reported fit parameters are from the F200W filter.}
\label{tab:fitparams}
\centering
\begin{tabular}{lll}
	    \hline
	    \hline
		Parameter & LMFit+Powell & Nested \\
		\hline
		  $r_0$ (AU) & 71.6 & $64.7_{-6.2}^{+6.5}$ \\
$\theta_{inc}$ (deg) & 72.0 & $70.1_{-3.3}^{+2.5}$ \\
$\theta_{PA}$ (deg) & -133.2 & $-132.9_{-2.1}^{+2.1}$ \\
$\alpha_{in}$ & 12.0 & $63.4_{-31.9}^{+23.7}$ \\
$\alpha_{out}$ & -4.6 & $-4.4_{-1.0}^{+0.6}$ \\
$h_0$ & 0.04 & $0.1_{-0.05}^{+0.11}$ \\
$g_1$ & 0.8 & $0.8_{-0.1}^{+0.1}$ \\
$g_2$ & -0.8 & $-0.9_{-0.1}^{+0.1}$ \\
$w_1$ & 0.9 & $0.9_{-0.1}^{+0.1}$ \\
$F_{max}$ (MJy/sr) & 39.2 & $44.5_{-18.7}^{+26.7}$ \\
		\hline
		\hline
	\end{tabular}
\end{table}


Parameter values are generally in agreement between the two fitting methods. However, some parameters were not well constrained by the nested sampling routine ($\alpha_{in}$, $h_0$). This may be due to degraded transmission towards the center of the coronagraph or due to the fact that the parameters are correlated. The orientation of the disk may also contribute to the difficulty in constraining certain parameters. For these parameters, the forward modeling best-fit values may fall towards the boundaries of or outside the uncertainties of the nested sampling values. Fig. \ref{fig:corner} in Appendix \ref{app:A} shows the nested sampling posterior distributions for each parameter.


\section{Search for Companions}
\label{sec:comp}

In keeping with the stated goal of the GO 4050 survey, which is to search for companions down to sub-Jupiter masses around host stars in the TW Hydrae star-forming region, we conducted an extensive search for companions in the TWA 20 system. In the final processed disk images, one candidate is identified in the F200W image by visual inspection that can also clearly be seen in the F444W image. 

\begin{figure*}
    \centering
    \includegraphics[width=\textwidth]{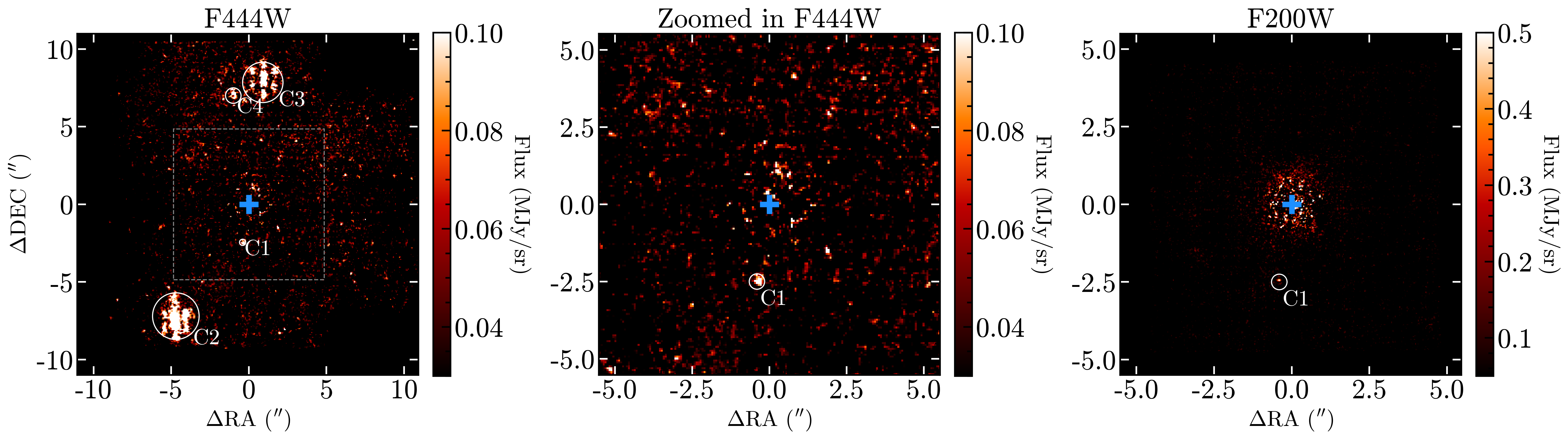}
    \caption{\label{fig:three200wcands} \textit{Left:} Disk-subtracted RDI image of the TWA 20 system in the F444W filter. Each off-axis point source is identified by a circle and a label. \textit{Middle:} Disk-subtracted RDI image of the TWA 20 system in the F444W filter zoomed in to the F200W FOV. \textit{Right:} Disk-subtracted RDI image of the TWA 20 system in the F200W filter.} 
\end{figure*}

In preparation for the companion search, we subtracted the disk model from the data and re-ran \texttt{spaceKLIP} image processing and then \texttt{spaceKLIP} RDI on the disk-subtracted data. This way, point sources that may have been hidden under the disk signal become more apparent. All off-axis points sources in the reference images for this stage were masked so as to not introduce spurious candidates to the science images. Fig. \ref{fig:three200wcands} depicts the disk-subtracted images in the F200W and F444W filters, where we highlight the locations of all companion candidates. In the F444W image, there is a bright cluster of pixels to the north of the disk ($\Delta$RA $\sim0.0$, $\Delta$DEC $\sim1.0$). By visual inspection at various stages of the data reduction process, we verified that these are not bad pixels. They correspond roughly to the position of a potential asymmetry in the disk seen in the F200W filter (see Fig. \ref{fig:modelsandresids} Data and Residuals for this filter at the stated $\Delta$RA and $\Delta$DEC). However, they do not resemble a planetary candidate PSF. This potential asymmetry is analyzed further in Section \ref{sec:disc}.

We then vetted the candidates with a variety of methods to determine if they are likely companions, background sources, or stellar speckles. For each candidate, we ran a PSF modeling routine in both filters. To do so, we used \texttt{spaceKLIP}'s \texttt{extract\_companions} function, which performs an MCMC fit of a model point source to the companion candidate by exploring right ascension (RA), declination (DEC), and flux parameter space with an initial guess for each parameter, plus the stellar age \citep[estimated to be 10 Myr based on the typical age of stars in the TW Hydrae region]{luhman2023census} and distance \citep[80.2 pc]{gaia2021} as input. Typically, an extended source such as a background galaxy or a data reduction artifact will not be fit well with a point source model. Such sources can often be ruled out as planets by visual inspection; we identify C2 as an extended source, and the various unlabeled bright spots in Fig. \ref{fig:three200wcands} as artifacts. However, candidates that are fit well by the model PSF may still be compact background galaxies or stars. To determine whether the source seen in both the F200W and F444W images (C1) is a planet, we compare its magnitudes in the F444W and F200W filters. If its color (F200W$-$F444W) and absolute magnitude in F444W follow the trend characteristic of observed giant planets \citep{saumon2008evolution, phillips2020new}, the candidate qualifies for further inspection. If it clearly does not follow this trend, C1 can be excluded as a potential planet. The models we used to derive the trend \citep[\texttt{BEX}, \texttt{ATMO}]{linder2019evolutionary,phillips2020new} compute the flux entering and exiting a planetary atmosphere (which can be used to calculate magnitude) and make several simplifying assumptions; models assume chemical equilibrium, no clouds, radiative-convective equilibrium, and 1D atmospheric structure (properties only vary vertically). Such assumptions are necessary to reduce computational complexity, but can reduce overall model accuracy.

\begin{figure}
    \centering
    \includegraphics[width=\linewidth]{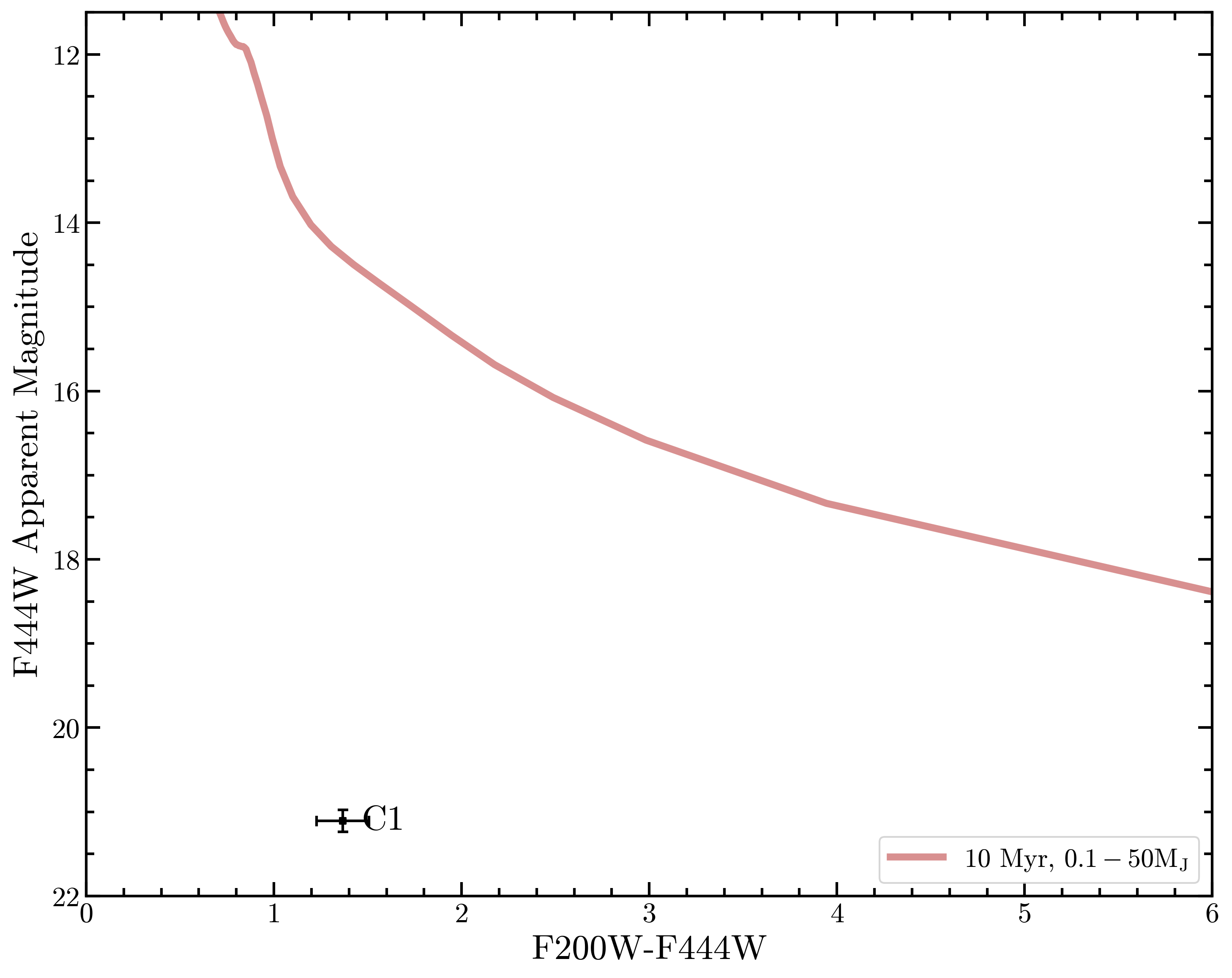}
    \caption{Color-magnitude diagram depicting the off-axis point source identified in both filters as well as the expected color-magnitude relationship for Jupiter-mass planets of a similar age to the TW Hydrae association (computed with \texttt{SPECIES} with planetary isochrone data from \citep{linder2019evolutionary,phillips2020new}). The point source clearly does not follow this trend, indicating it is likely a background source.}
    \label{fig:rejects}
\end{figure}

Fig. \ref{fig:rejects} demonstrates the position of C1 (the candidate in the FOV of both the F200W and F444W images) on a color-magnitude diagram. In this plot, we also include a curve indicating the expected color magnitude trend for giant planets of a similar age to the TWA 20 system. We compute this curve with the \texttt{SPECIES} Python software \citep{stolker2020miracles}, using planetary isochrone data from \cite{linder2019evolutionary} (\texttt{BEX}) and \cite{phillips2020new} (\texttt{ATMO}) and an age of 10 Myr as input into the \texttt{get\_color\_magnitude} function. It is clear that the candidate does not follow the color-magnitude trend for giant planets represented by the solid line, indicating it is likely to be a background source.

Fig. \ref{fig:three200wcands} also highlights the locations of several additional companion candidates in the F444W filter. Planetary candidates may be detected only in F444W and not in F200W; however, these objects are outside of the FOV of the F200W filter, so we report their fitted magnitudes but do not investigate them further as planetary candidates. C2 appears to be an extended source. Any other notably bright spots that are not labeled in Fig. \ref{fig:three200wcands} are likely data reduction artifacts. Table \ref{tab:candmags} summarizes the apparent magnitudes derived for all off-axis sources across the two filters.

\begin{table}
\caption{Calculated apparent magnitude for each of the off-axis sources in the TWA 20 science images. Off-axis sources with only F444W magnitude reported are outside the FOV of the F200W filter.}
\label{tab:candmags}
\centering

\begin{tabular}{llll}
	    \hline
	    \hline
		Candidate & F200W & F444W & Separation ($\arcsec$) \\
		\hline
		  C1 & 21.07$\pm$0.13 & 21.11$\pm$0.13 & 2.5 \\
            C2 &  & 18.61$\pm$0.06 & 8.9  \\
            C3 &  & 18.07$\pm$0.05 & 8.1  \\
            C4 &  & 19.79$\pm$0.06 & 7.0  \\
		\hline
		\hline
	\end{tabular}
\end{table}

\begin{figure*}
    \centering
    \includegraphics[width=\textwidth]{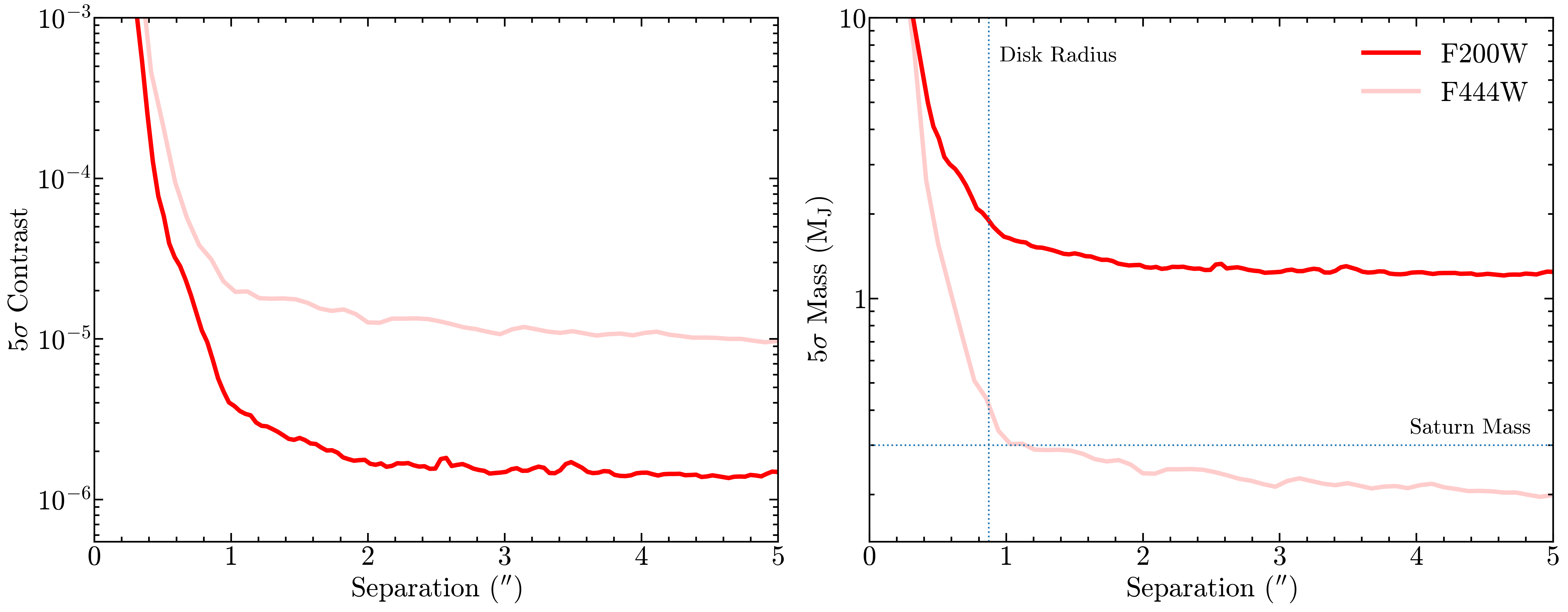}
    \caption{\label{fig:contcurves} Calibrated contrast curves (\textit{left}) and mass sensitivity limits (\textit{right}) for the TWA 20 system in each filter as a function of projected separation. At separations of 1\arcsec, equivalent to 80 AU, 2$\times10^{-5}$ (3$\times10^{-6}$) contrasts are achievable with JWST NIRCam in the F444W (F200W) filter, corresponding to planetary masses of 0.3$M_J$ (2$M_J$). The F444W filter is sensitive to lower planetary masses than the F200W filter despite achieving worse contrasts due to the fact that planets are often much brighter at the wavelengths of the F444W filter. The dashed vertical line corresponds to the disk peak radius ($r_0$). The dashed horizontal line corresponds to the mass of Saturn.} 
\end{figure*}

To inform future companion searches in the TWA 20 system, we also computed contrast curves from the images in each filter. These curves are shown in the left-hand plot of Fig. \ref{fig:contcurves}.

For this calculation, we again use \texttt{spaceKLIP}, which determines contrast in several steps. First, \texttt{spaceKLIP} estimates the stellar flux in the given filter by modeling the stellar PSF. The science image is then normalized to the stellar flux. Next, \texttt{spaceKLIP} measures the science image noise by determining the standard deviation of the pixels in radial bins at various distances from the host star, accounting for small sample statistics \citep{mawet2014fundamental}. To model the stellar PSF, we input the stellar spectral type and stellar photometric data from the \textit{Spitzer}, \textit{ALLWISE}, SDSS, SkyMapper, Gaia, and 2MASS surveys \citep{spitzerdata,allwisedata,sdssdata,skymapperdata,gaia2021,2massdata} accessed with the VizieR online photometry viewer tool \citep{vizier}. Third, \texttt{spaceKLIP} accounts for the variable throughput of the coronagraphic mask by scaling the magnitude of each pixel in the science image accordingly. The raw contrast is then the threshold for a planet detection that is $5-\sigma$ above the image noise in a given radial bin. For the F200W filter, contrasts beyond 1" separation are below $4\times10^{-6}$; for the F444W filter, contrasts beyond 1" separation are below $3\times10^{-5}$.


From the contrast curves, we also estimated the sensitivity of JWST NIRCam to different planetary masses in the TWA 20 system. To do so, we used the \texttt{SPECIES} Python software \citep{stolker2020miracles}to calculate mass as a function of projected separation with the \texttt{contrast\_to\_mass} function using planetary isochrone data from \cite{linder2019evolutionary} and \cite{phillips2020new}, the estimated age of the star, and the distance to the star. The planetary mass sensitivity curves can be seen in the right-hand plot of Fig. \ref{fig:contcurves}. For the F200W filter, NIRCam is sensitive to masses below 2$M_J$ beyond 1" separation; for the F444W filter, NIRCam is sensitive to masses below 0.3$M_J$ beyond 1" separation. From the distance to TWA 20, this separation corresponds to $\sim$80 AU. Regarding these sensitivities, it must be noted that the \cite{linder2019evolutionary} and \cite{phillips2020new} models assume chemical equilibrium and no clouds; planets have been shown to be fainter than these models predict \citep[e.g. TWA 7][]{lagrange2025evidence}. Uncorrected stellar speckles cause substantial degradation of the contrast and thus the mass sensitivity of the inner regions of the system.

\section{Discussion} \label{sec:disc}

\begin{figure}
    \centering
    \includegraphics[width=\linewidth]{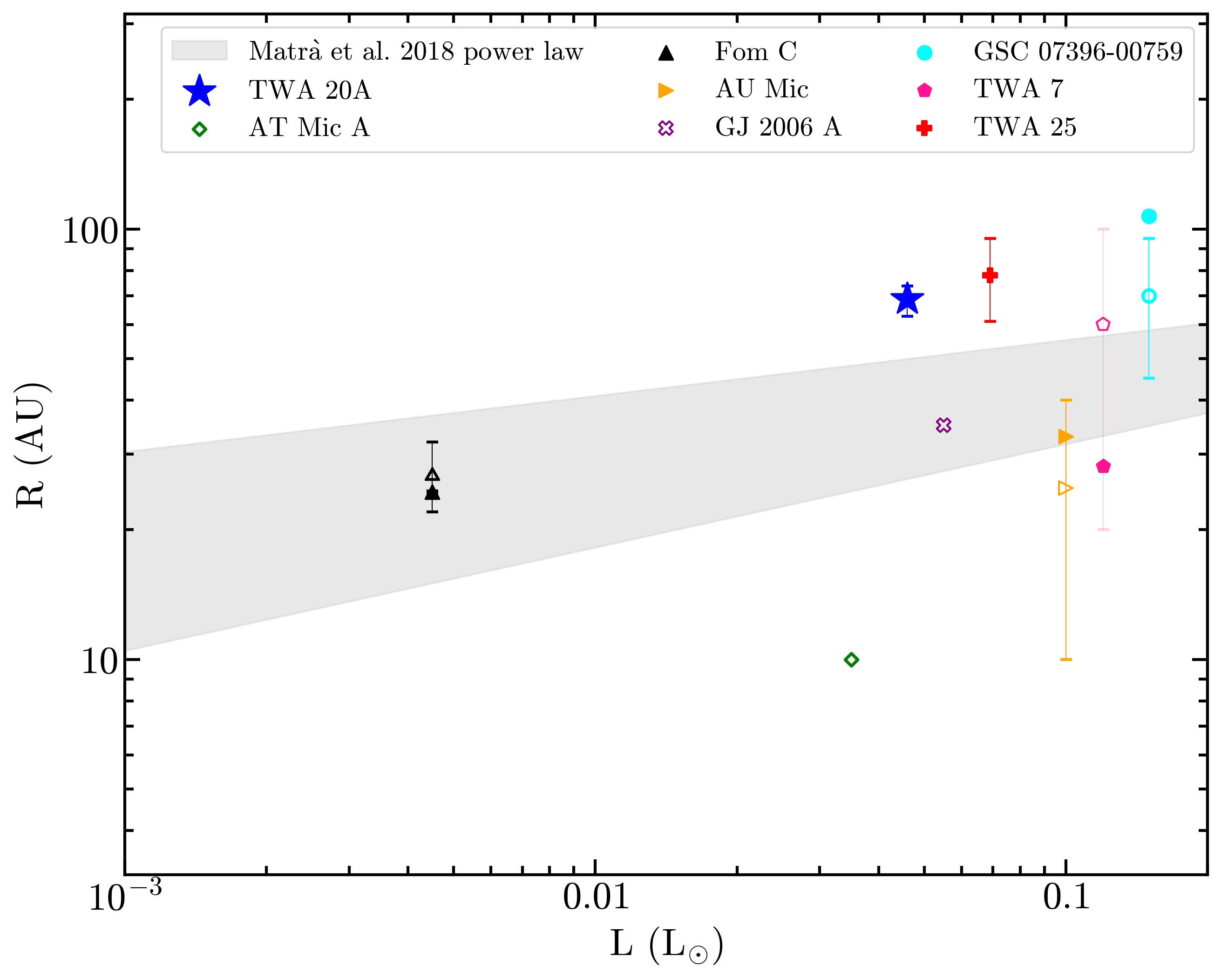}
    \caption{\label{fig:Mdiskrad} Debris disk radius as a function of host-star luminosity for resolved M dwarf disks. Scattered light radii are shown as opaque/filled markers; thermal emission radii are shown as open/unfilled markers. Both radius measurements are reported for disks with both scattered light and thermal emission data. Disk radii and stellar luminosity were taken from \cite{matra2025resolved}, \cite{choquet2016first}, \cite{lawson2023jwst}, \cite{lawson2024jwst}, \cite{adam2021characterizing}, and \cite{cronin2023alma}. Only the inner most ring is included for TWA 7 for visual clarity; the additional rings extend well beyond 100 AU Luminosity for TWA 20 was computed from the TWA 20 photometric data. The shaded area represents the power law relationship between R and L$_*$ determined by \cite{matra2018empirical, matra2025resolved}.} 
\end{figure}

The TWA 20 debris disk is the 6$^{th}$ debris disk detection in scattered light and the eighth resolved disk (in scattered light or thermal emission) around an M dwarf. The small library of M dwarf disks exhibit a variety of geometric properties with substantial variation in radius and brightness. Among these detections, the TWA 20 system has a best-fit radius ($r_0\sim65$AU) that is larger than that of five of the eight imaged M dwarf disks shown in Fig. \ref{fig:Mdiskrad}. TWA 20 is the largest disk orbiting a host star with luminosity $\leq0.06L_{\odot}$, though TWA 25A has a similar radius and only a slightly larger stellar luminosity. In Fig. \ref{fig:Mdiskrad}, we show the $1-\sigma$ bounds of the R-L$_*$ power-law relationship for debris disks derived by \cite{matra2018empirical, matra2025resolved}, along with data from the resolved M dwarf disks. Thermal emission data for Fom C, AU Mic, GSC 07396-00759, and TWA 7 were adopted from \cite{matra2018empirical}. Thermal emission data for GJ 2006 A and AT Mic A were adopted from \cite{cronin2023alma}. Scattered light data for TWA 25, Fom C, AU Mic, GSC 07396-00759, and TWA 7 were adopted from \cite{choquet2016first}, \cite{lawson2024jwst}, \cite{lawson2023jwst}, \cite{adam2021characterizing} and \cite{crotts2025follow}, respectively. The relationship from \citet{matra2018empirical, matra2025resolved} was determined by analyzing 26 debris disks of various spectral types in thermal emission; disks around M dwarfs are expected to follow the power-law trend similarly well to disks around host stars of other spectral types. Compared to the properties of the other disks and the power law in Fig. \ref{fig:Mdiskrad}, TWA 20 appears to be relatively typical.

\begin{figure*}
    \centering
    \includegraphics[width=\textwidth]{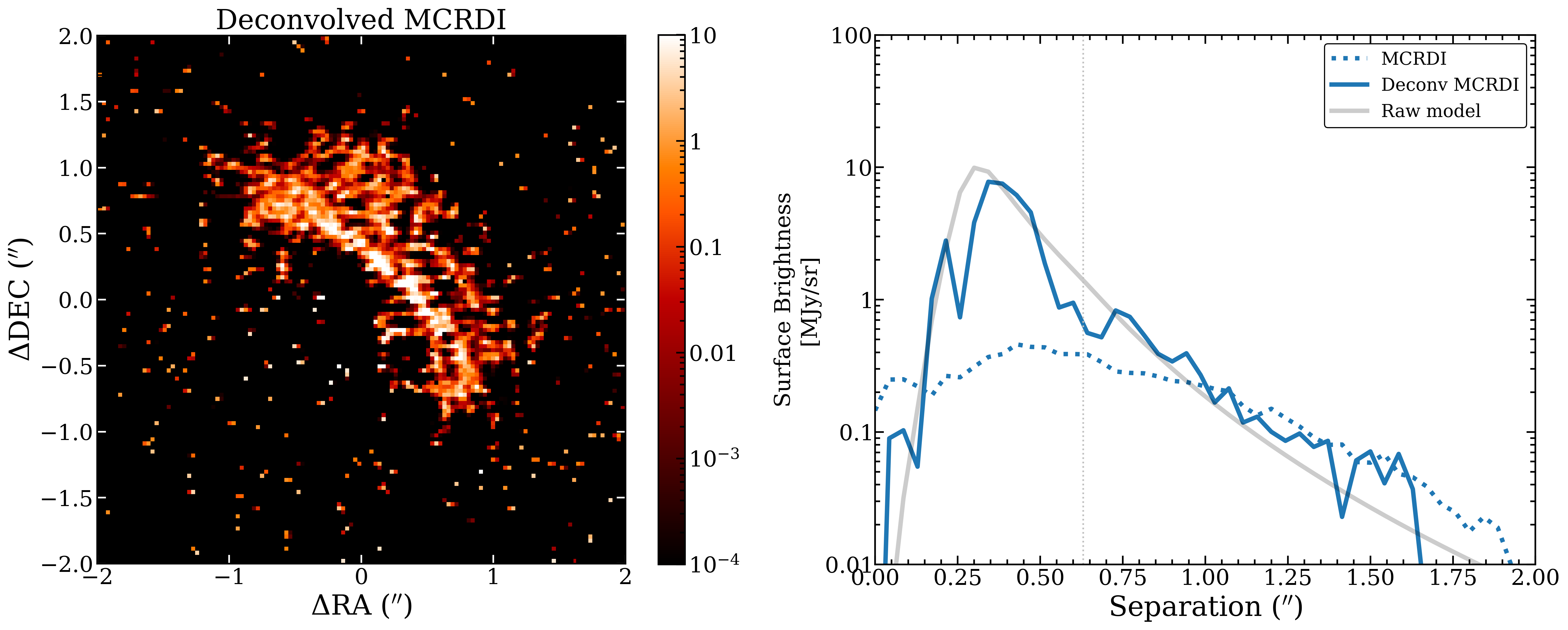}
    \caption{\label{fig:decon} \textit{Left: } De-convolved image of the TWA 20 debris disk in the F200W filter obtained by running a deconvolution routine on the MCRDI result. The image is rotated North up. \textit{Right:} Surface brightness (radial profiles) of the de-convolved disk, the raw disk model (analytical model from the fitting procedure), and the convolved disk as a function of separation from the host star.} 
\end{figure*}

The TWA 20 disk average surface brightness is depicted by the brightness profile of the de-convolved disk image in Fig. \ref{fig:decon}, constructed by taking the mean of each of a series of elliptical annuli centered at $\Delta{RA}, \Delta{DEC} = (0,0)$ with increasing semimajor and semiminor axes. This value may differ substantially from the measured $F_{max}$, as $F_{max}$ is the global peak surface brightness while here we report the radial average surface brightness. We extracted surface brightness using a \texttt{Winnie} deconvolution routine to avoid underestimation of the brightness due to the effects of the JWST PSF. The peak of $\sim10$ MJy/sr in the F200W filter is comparable to that of other M dwarf debris disks \citep{choquet2016first,crotts2025follow}, though drawing any physical insights from this is difficult due to the different inclination of each disk. The peak brightness of TWA 20 in the F444W filter, $\sim0.1$ MJy/sr, is similar to that of Fomalhaut C \citep[$\sim0.05$ MJy/sr]{lawson2024jwst} in this filter; both sources have a similar spectral type \citep[Fom C is M4V]{davison20153d} and low brightness in this filter compared to other earlier M dwarf disks \citep{lawson2023jwst,crotts2025follow}, which emphasizes the difficulty in detecting mid (and later) M dwarf disks in the F444W filter with current instrumentation. 

We note that in the de-convolved TWA 20 image as well as the convolved RDI images, there is an additional bright feature towards the ``top" of the disk ($\Delta$RA $\sim0.0\arcsec$, $\Delta$\textbf{DEC} $\sim1.0\arcsec$). It is unclear if this feature is part of the stellar system or is additional light from some background source. If the structure is not part of the system, it may bias our reported surface brightness as well as our reported disk properties. If it is part of the system, then it contributes to a growing list of disks with non-axisymmetric features, which includes other M dwarf disks such as AU Mic and TWA 7 \citep{fitzgerald2007microscopii,crotts2025follow}. The color between the feature and the main disk---measured by computing the average flux in a square region centered at the brightest point in the disk and around the feature, converting this flux to a magnitude, and taking the magnitude difference between the disk and the feature---is relatively small in both filters (0.2$\pm1.2$ in F200W and 0.08$\pm0.2$ in F444W). This may suggest that the feature is indeed a part of the disk. As part of the disk, this asymmetric structure may not be fit well by our disk model, which assumes axisymmetry. Further determination of the origin of this feature, including further disk modeling with an eccentric model, is left for future work. The color of the disk itself, computed as the difference between the disk magnitude in each filter subtracted from the difference between the stellar magnitude in each filter, is $0.2\pm0.5$. This is consistent with the neutral color observed for other disks at 2\textmu{m} and beyond \citep{crotts2024uniform}.

\begin{figure}
    \centering
    \includegraphics[width=\linewidth]{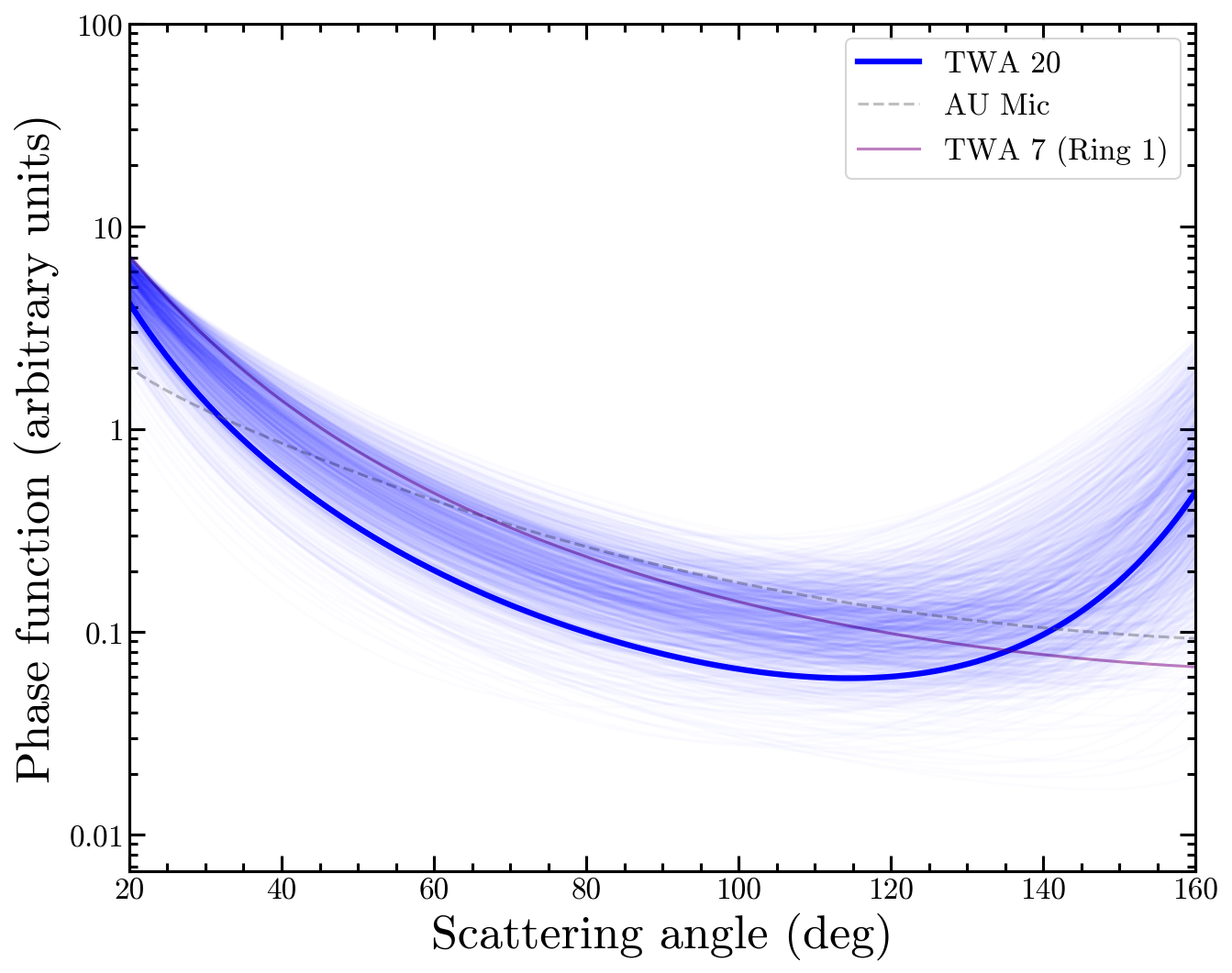}
    \caption{\label{fig:spf} The HG SPF as a function of scattering angle for TWA 20, AU Mic \citep{lawson2023jwst}, and TWA 7 \citep{crotts2025follow}, based on disk models fit to scattered light observations with JWST/NIRCam. In addition to the SPF constructed with the best-fit values of $g_1$, $g_2$, and $w_1$ for TWA 20 (thick blue line), SPF's based on 1000 random draws from the posterior distributions of $g_1$, $g_2$, and $w_1$ are plotted to demonstrate the spread in possible SPF structure. In most of these draws, a notable backward-scattering peak is visible in addition to the forward-scattering peak characteristic of the SPF of all three plotted disks.} 
\end{figure}

Other properties to consider are the radial density power law exponents. Since we only obtained tight constraints on $\alpha_{out}$ (and not $\alpha_{in}$) with our modeling procedure, we only compare our value for $\alpha_{out}$ with that of other disks. Our value of $\alpha_{out} = -4.42$ is similar to values reported for AU Mic \citep{lawson2023jwst} and for the brightest ring of TWA 7 \citep{crotts2025follow}, but closer to zero (i.e. a shallower disk inner-edge) than values reported for GSC 07396-00759 \citep{cronin2022alma}. We also do not get tight constraints on $h_0$, but our best-fit value roughly matches that of TWA 7 \citep{crotts2025follow} and AU Mic \citep{lawson2023jwst}. The SPF parameters, $g_1$, $g_2$, and $w_1$, are somewhat constrained though there is some degeneracy between $g_2$ and $w_1$. Fig. \ref{fig:spf} visualizes the SPF of TWA 20 based on 1000 random draws from the posterior distributions of $g_1$, $g_2$, and $w_1$. Though the measured values are consistent with more forward-scattering than backward-scattering, there is a notable backward-scattering peak that can be seen in Fig. \ref{fig:spf}. This peak is not identified in JWST scattered-light observations of other M dwarf disks (AU Mic and TWA 7) based on their reported best-fit values for $g_1$, $g_2$, and $w_1$ \citep{lawson2023jwst, crotts2025follow}. That said, it must be stated that the backside of the disk is masked by the coronagraph (see the convolved vs. raw model in Fig. \ref{fig:models}), and so most back-scattering angles are poorly probed by the data. The peak thus may be created by the forced shape of the SPF based on the structure of the scattering angles that are accessible. Determination of the disk grain structure and composition and how it relates to the measured SPF is left to future work. Follow-up observations may help to further constrain the various model parameters and thus facilitate further comparison between TWA 20 and the other M dwarf disks. Such comparisons will help us improve our understanding of disk formation processes and ultimately determine primary mechanisms for disk dynamics and morphologies around low mass stars.

As evidenced by TWA 20's low surface brightness in F444W and the fact that TWA 20 has not been detected previously, TWA 20 is a difficult disk to detect for telescopes other than JWST. As a mid-M disk, detection in scattered light from the ground is likely quite difficult due to the atmospheric background noise in the near-infrared and at longer wavelengths. The host star is cool and red, and as such the disk signal may be weak in the visible \citep[because the disk is scattering more strongly in the NIR]{lawson2024jwst} and the thermal IR (because the disk is cold). It is possible that this disk could be detected with HST NICMOS at NIR wavelengths, but at visible wavelengths, TWA 20 may not scatter strongly enough to be detectable. As mentioned previously, no IR excess has been detected for this system, but it is difficult to assess whether follow-up observations at mm wavelengths (e.g. with ALMA) would yield a detection due to uncertainty about the grain composition and grain size distribution of the disk. 

\begin{figure}
    \centering
    \includegraphics[width=\linewidth]{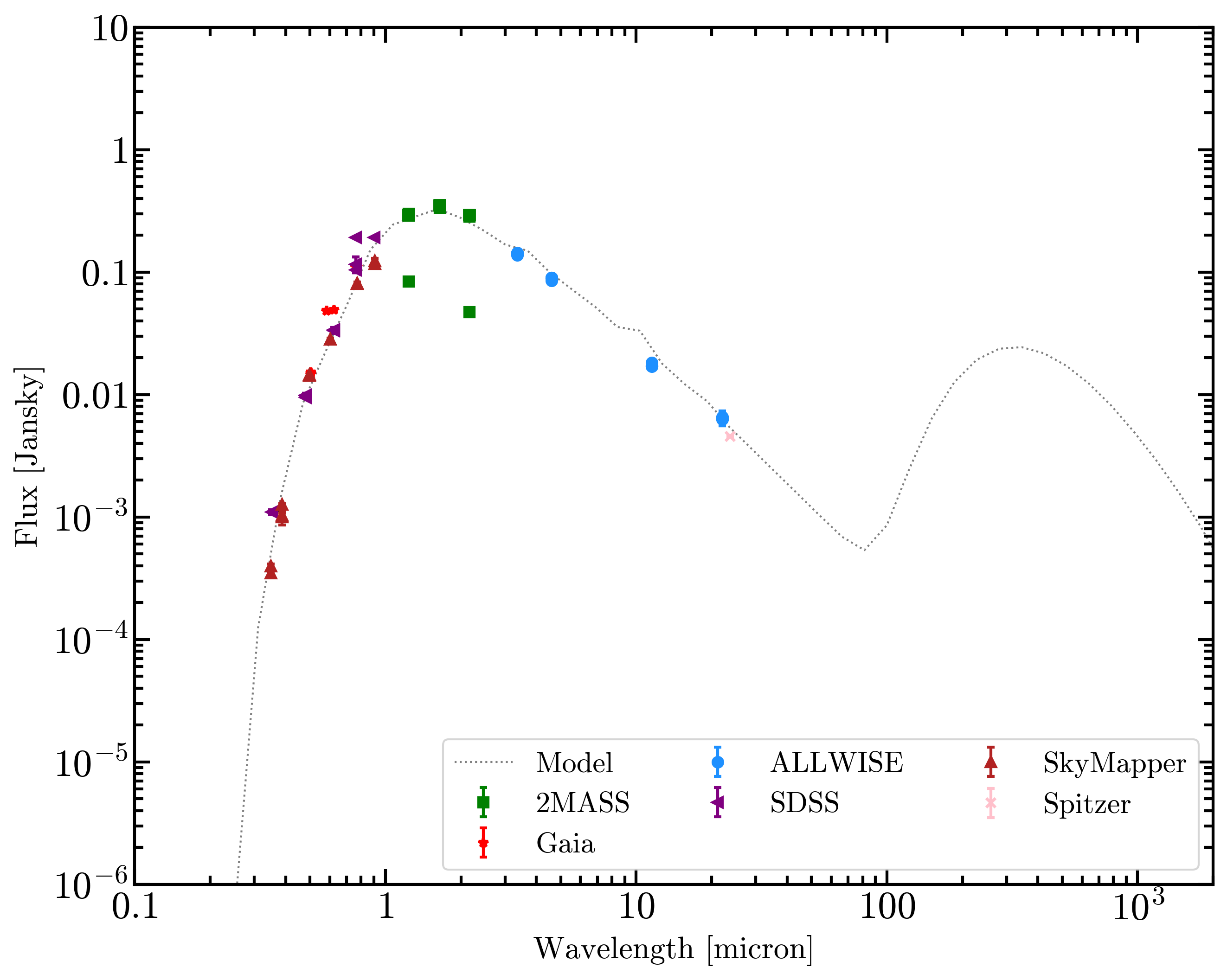}
    \caption{\label{fig:sed} TWA 20 photometry data from \citet{allwisedata,gaia2021,2massdata,skymapperdata,spitzerdata,sdssdata} and our \texttt{MCFOST} model stellar + disk SED. For this target, there are no observations at longer wavelengths to validate the potential excess beyond 100 \textmu{m}.}
\end{figure}

We use the known morphological parameters from the model fitting in this work to model the SED with the \texttt{MCFOST} radiative transfer simulation software \citep{pinte2022mcfost}. We assume typical astrosilicate dust, and scale the dust mass such that the model disk brightness matches that of the data. From these SED models, we find that our scattered light detection is consistent with a disk having an excess at wavelengths beyond $\sim$100 \textmu{m}. MCFOST returns the stellar and disk SEDs, which we present alongside the photometry as a qualitative comparison, as shown in Fig. \ref{fig:sed}. There are no observations at the wavelengths where the IR excess may be detectable that should have been sensitive to this excess. Though TWA 20 was never observed with the \textit{Herschel} Photodetector Array Camera and Spectrometer (PACS), the sensitivity limits of this instrument at wavelengths where the excess may be present ($\sim2$ mJy at 100 \textmu{m} and $\sim4.4$ mJy at 160 \textmu{m}) \citep{lutz2011pacs} do not suggest that the excess would be clearly detected (our SED model has values of $\sim1$ mJy at 100 \textmu{m} and $\sim3$ mJy at 160 \textmu{m}). Integrating the stellar SED and the disk model SED and taking the ratio of model luminosity to stellar luminosity, we find a fractional luminosity of $\sim1.6\times10^{-4}$. However, we do not pursue more precise quantitative thermal emission estimates due to the large number of assumptions inherent in this type of modeling. The scattered light disk detection is in only one filter; such a detection provides poor constraints on grain structure and composition. A range of different dust characteristics could produce the scattered light measurements we report but would result in a different SED.

TWA 20 is one of six M-dwarf debris disks that have been imaged in scattered light, and among these is one of two that show no measured IR excess \citep{sissa2018new}. Because M-dwarfs have low luminosity in comparison to stars of other spectral types, the disk material may be cooler and thus less likely to be detected as an IR excess \citep{matthews2007mass}; there are various additional reasons why M dwarf disks have been difficult to detect in IR photometric surveys explored in other works (e.g. \citep{luppe2020MNRAS}). Thus, despite the fact that IR excess is a common signature of debris disks around stars of other spectral types \citep{thureau2014unbiased,chen2020spitzer}, it is possible that TWA 20 has no detectable infrared excess. However, the large number of assumptions in this model prevent any definitive conclusions about grain size and composition. As more M-dwarf disks are detected and characterized, a clearer picture may emerge about their grain properties and the physical processes that create them.

\section{Conclusion} \label{sec:conc}
In this work, we report on the imaging discovery of a debris disk in scattered light around the TWA 20 host star. The images were obtained with JWST/NIRCam as part of the GO 4050 survey and processed using RDI and MCRDI. By fitting a model to the disk in the F200W image (where the detection is high SNR), we determine a disk radius of 64.7$_{-6.2}^{+6.5}$ AU, an inclination of 70.1$^{\circ}$$_{-3.3}^{+2.5}$, and a PA of -132.9$^{\circ}$$_{-2.1}^{+2.1}$. We find that compared to other known M star disks, TWA 20 has a comparable radius and brightness. We report no companion detections in the system, and determine mass-sensitivity limits of 0.3$M_{J}$ at 80 AU in the F444W filter and 1.6$M_{J}$ at 80 AU in the F200W filter for JWST/NIRCam. Based on the F444W mass-sensitivity curve and the measured disk radius, we can likely rule out a Jupiter mass perturber at separations wider than 48 AU. Follow-up observations may help to further constrain disk geometry, enable dust grain characterization, refine the reported mass sensitivity limits, and/or detect new companions. 

The presented results highlight JWST's unique sensitivity to scattered light debris disks around M stars, which have been historically difficult to detect. Given the abundance of these stars in the Milky Way galaxy, our findings are an encouraging addition to the catalog of M dwarf disk discoveries, which will continue to grow and enhance our understanding of planetary systems around low mass stars.

\begin{acknowledgments}
\section{Acknowledgements}
\end{acknowledgments}
This work is based on observations with the NASA/ESA/CSA JWST, obtained at the Space Telescope Science Institute, which is operated by AURA, Inc., under NASA contract NAS 5-03127. These observations are associated with the JWST program 4050 (PI: A. Carter). The JWST data presented in this article were obtained from the Mikulski Archive for Space Telescopes (MAST) at the Space Telescope Science Institute. Support for program 4050 was provided by NASA through a grant from the Space Telescope Science Institute, which is operated by the Association of Universities for Research in Astronomy, Inc., under NASA contract NAS 5-03127. This work benefited from the 2024 and 2025 Exoplanet Summer Program in the Other Worlds Laboratory (OWL) at the University of California, Santa Cruz, a program funded by the Heising-Simons Foundation.

S.P. acknowledges the significant harm caused to members of the LGBTQIA+ community in the Department of State and NASA, while under the leadership of James Webb as Under Secretary of State and NASA Administrator, respectively.

S.P. and M.M.B. are supported by Space Telescope Science Institute under Grant No.~JWST-GO-04050.015-A.

A.S. is supported by the National Science Foundation Graduate Research Fellowship under Grant No.~2139433.

S.M. acknowledges funding by the Royal Society through a Royal Society University Research Fellowship (URF-R1-221669) and the European Union through the FEED ERC project (grant number 101162711). Views and opinions expressed are however those of the author(s) only and do not necessarily reflect those of the European Union or the European Research Council Executive Agency. Neither the European Union nor the granting authority can be held responsible for them.

This material is based upon work supported by the National Science Foundation Astronomy \& Astrophysics Postdoctoral Fellowship Award No. 2401654 for author B.L. Any opinions, findings, and conclusions or recommendations expressed in this material are those of the authors(s) and do not necessarily reflect the views of the National Science Foundation.

R.B.W. is supported by a Royal Society grant (RF-ERE-221025).

The JWST data used in this paper can be found in MAST: \dataset[10.17909/s1ea-h159]{http://dx.doi.org/10.17909/s1ea-h159}.

\vspace{5mm}
\facilities{JWST(NIRCam)}


\software{astropy \citep{astropy:2013,astropy:2018,astropy:2022}, \texttt{Winnie} \citep{Lawson22,lawson2023jwst}, \texttt{ibadpx} \citep{ibadpx}, \texttt{STPSF} \citep{perrin2014updated}, \texttt{LMFit} \citep{lmfit25}, \texttt{DYNESTY} \citep{2020MNRAS.493.3132S,sergey_koposov_2024_12537467}, \texttt{SPECIES} \citep{stolker2020miracles}, \texttt{MCFOST} \citep{pinte2022mcfost}
          }



\appendix
\section{Additional Figures}
\label{app:A}
\begin{figure*}
    \centering
    \includegraphics[width=\textwidth]{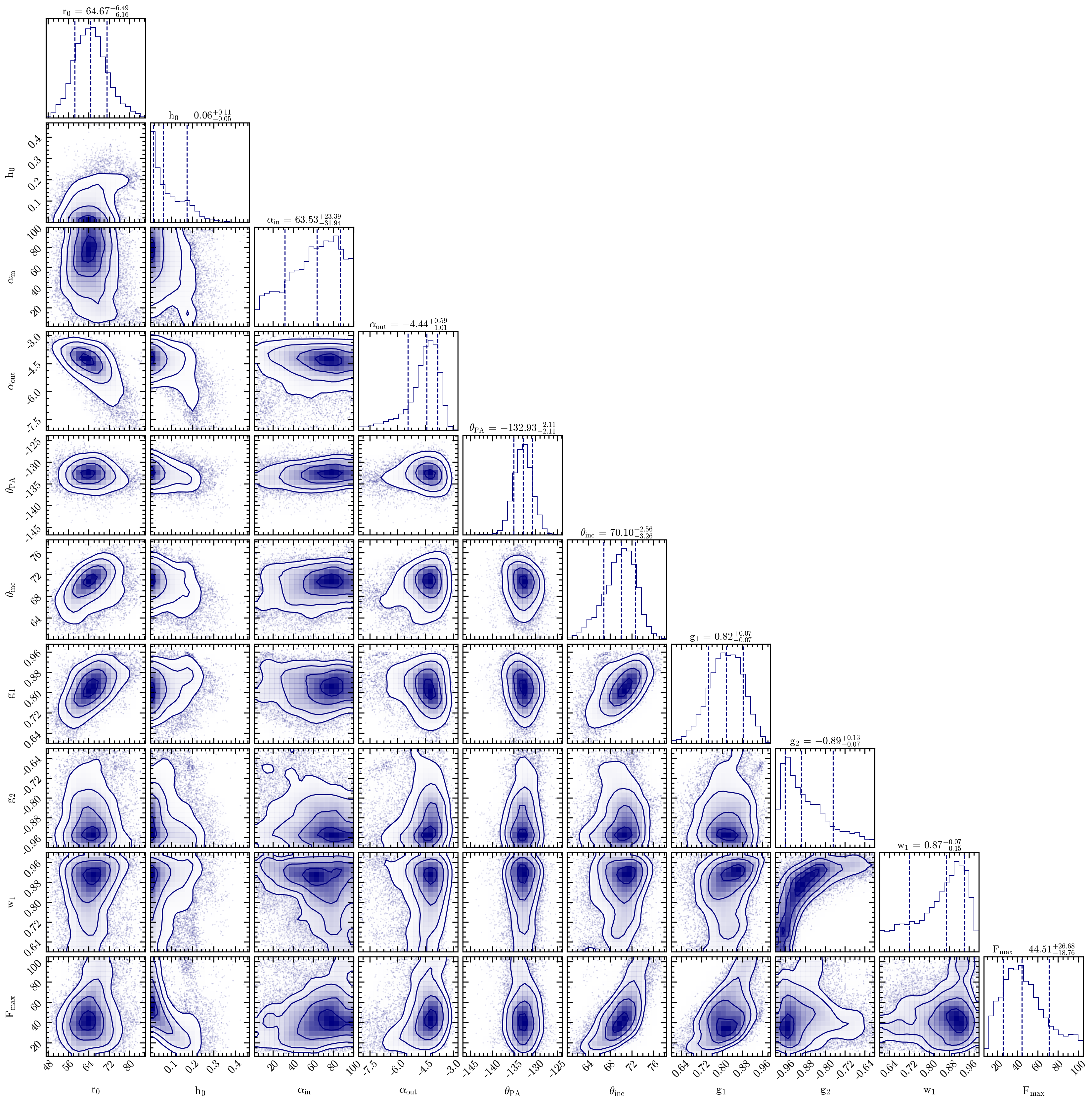}
    \caption{\label{fig:corner} \textit{Left: } Posterior distributions from the nested sampling disk fitting routine. Some parameters, such as $r_0$, $\alpha_{out}$, $\theta_{inc}$, and $\theta_{PA}$, are well constrained; other parameters, such as $\alpha_{in}$ and $F_s$ are not. Follow-up observations may improve our understanding of the structure of the TWA 20 disk.} 
\end{figure*}

\bibliography{sample631}{}
\bibliographystyle{aasjournal}



\end{document}